\documentclass[aps,prd,twocolumn,nofootinbib,superscriptaddress,floatfix]{revtex4-2}

\usepackage{amsmath}
\usepackage{amssymb}
\usepackage{bm}
\usepackage{graphicx}
\usepackage{booktabs}
\usepackage{dcolumn}
\usepackage{xcolor}
\usepackage[colorlinks=true,citecolor=blue,linkcolor=blue,urlcolor=blue]{hyperref}
\usepackage[normalem]{ulem}

\graphicspath{{figures/}}

\newcommand{\Msun}{M_{\odot}}

\newcommand{\Ctwelve}{\mathrm{C}^{12}}
\newcommand{\Ecal}{\mathcal{E}}
\newcommand{\Gcal}{\mathcal{G}}
\newcommand{\MADM}{M_{\rm ADM}}
\newcommand{\Mint}{M_{\rm int}}
\newcommand{\Mgeom}{M_{\rm g}}
\newcommand{\Rareal}{R_{\rm areal}}

\newcommand{\lc}{\mathring{\nabla}}
\newcommand{\GEin}{\mathring{G}}

\begin{document}

\title{Bayesian constraints on quadratic $f(Q)$ gravity from the ADM mass of white dwarfs}

\author{Edson Otoniel}
\affiliation{Universidade Federal do Cariri (UFCA), Instituto de Forma\c{c}\~ao de Educadores, Brejo Santo, CE 63260-000, Brazil}
\email{edson.otoniel91@ufca.edu.br}
\author{Lucas A. Souza}
\affiliation{Independent Researcher, Minas Gerais, Brazil}
\email{lucasufsj@gmail.com}
\author{Jonathan A. Rebou\c{c}as}
\email{jalvesreboucas@ifce.edu.br }
\affiliation{Instituto Federal de Educa\c{c}\~ao, Ci\^encia e Tecnologia do Cear\'a (IFCE), Iguatu, CE, Brazil}
\affiliation{Universidade Estadual do Cear\'a (UECE), Faculdade de Educa\c{c}\~ao, Ci\^encias e Letras de Iguatu, Iguatu, CE 63500-000, Brazil}

\author{Juan M. Z. Pretel}
\email{juan04manuel91@gmail.com}
 \affiliation{Centro Brasileiro de Pesquisas F{\'i}sicas, Rua Dr.~Xavier Sigaud, 150 URCA, Rio de Janeiro CEP 22290-180, RJ, Brazil
}

\date{\today}

\begin{abstract}
We calculate static equilibrium sequences for white dwarfs (WDs) in the quadratic symmetric teleparallel model \(f(Q)=Q+\alpha Q^2\) and perform a Bayesian comparison of the negative coupling sector, general relativity (GR), and the positive coupling sector using mass and radius measurements. The stellar structure is obtained within a spherically symmetric geometry with a general areal radius function by imposing the affine field equation and selecting the regular analytic branch \(C=r e^{B/2}\). Each admissible interior solution is then matched to an asymptotically flat exterior spacetime, from which the ADM (Arnowitt–Deser–Misner) mass at spatial infinity is obtained. The Bayesian likelihood is constructed using this ADM mass, while the integrated energy density $M_{\rm int}$ is retained solely as a consistency diagnostic. The matter sector is described by a cold \(\Ctwelve\) equation of state (EoS) with a relativistic degenerate electron gas, nuclear rest energy, and Coulomb lattice contribution. The unmeasured central density is marginalized at fixed \(\alpha\) with a prior uniform in \(\log\rho_c\). The signed sequences cover \(10^{14}\leq|\alpha|/\mathrm{cm^2}\leq10^{21}\). Positive couplings shift the compact branch toward larger radii and lower masses, whereas negative couplings generate high mass extensions, including an admitted equilibrium configuration with \(M_{\rm ADM}=14.34\,M_\odot\). The reference inference uses Sirius B, QS Vir, V471 Tau B, ZTF J1901+1458, and LHS 4033. Within each continuous sign sector, the prior is uniform in \(\log_{10}(|\alpha|/\mathrm{cm^2})\). The conditional posterior medians are \(\alpha_-^{\rm med}=-1.12\times10^{16}\,\mathrm{cm^2}\) and \(\alpha_+^{\rm med}=1.85\times10^{17}\,\mathrm{cm^2}\). With equal prior weights for the negative, GR, and positive sectors, their posterior probabilities are \(0.228\), \(0.378\), and \(0.395\), respectively, with \(\ln B_{+/{\rm GR}}=0.043\). Therefore, the available data yield conditional constraints on the coupling magnitude within each sign sector and retain comparable integrated probability for the three gravitational descriptions, rather than a direct signature of modified gravity, highlighting that the ADM mass prescription and the admissible stellar domain must be properly defined before WD mass and radius observations can be employed to constrain the theory.

\end{abstract}

\maketitle

\section{Introduction}
\label{sec:introduction}

The mass and radius relation of WDs follows from a balance between gravity and electron degeneracy pressure. Chandrasekhar showed that the relativistic electron gas establishes a limiting mass for cold ideal WD configurations \cite{Chandrasekhar1931, Chandrasekhar1935}. More complete treatments include Coulomb corrections, finite nuclear mass contributions, composition-dependent reaction thresholds, and the thermodynamics of dense matter \cite{Salpeter1961, HamadaSalpeter1961, KoesterChanmugam1990, Chamel2013}. Consequently, deviations from the standard mass--radius relation may arise from either modified gravity or the adopted microphysical model. These considerations imply that WDs provide meaningful tests of gravity only once the underlying microphysics and the validity domain of the adopted EoS have been clearly established.

The observational status of the WD mass--radius plane has changed substantially with precise parallaxes and binary constraints. Gaia-based catalogues enlarged the available WD population and improved the statistical control of distances and photometry \cite{Tremblay2017, GentileFusillo2021}. Benchmark stellar systems, such as Sirius B, the eclipsing binaries QS Vir and V471 Tau, and several wide binaries, now provide direct or nearly direct determinations of stellar masses and radii. Combined with gravitational redshifts, atmosphere models, and flux calibration, these systems offer some of the most stringent observational tests of WD structure \cite{Bond2017, JoyceSirius2018, Parsons2016, Muirhead2022, Arseneau2024}. The resulting data set provides one of the most robust observational benchmarks currently available for confronting WD structure models and placing quantitative constraints on modified gravity theories.

Despite these observational advances, important astrophysical degeneracies remain. The inferred stellar radii remain sensitive to atmosphere models, temperature, envelope mass, chemical composition, and distance calibration. Likewise, theoretical WD models depend on the hydrogen or helium layer, Coulomb interaction terms, electro-capture thresholds, magnetic stresses, and rotation \cite{Romero2019,Caiazzo2021,Dahn2004,Otoniel2015Fermionic,Otoniel2017Rotation,Otoniel2019Magnetized,Otoniel2021CTCV,Sousa2024GW}. Consequently, meaningful constraints on gravity require disentangling gravitational effects from uncertainties associated with the adopted microphysics and the interpretation of WD observations.

Modified gravity theories can alter the structure of compact stars in ways that are partially degenerate with the properties of dense stellar matter. This issue is well known in curvature-based theories and in models with nonminimal matter couplings, where the same mass--radius curve may be reproduced through different combinations of the EoS, gravitational coupling, or exterior mass definition \cite{Carvalho2017, Astashenok2022PRD, Otoniel2026}. In fact, WDs have been investigated within alternative theories of gravity, including Eddington-inspired Born-Infeld gravity \cite{Pani2012, Banerjee2022}, Palatini $f(R)$ theory \cite{Sarmah2022PRD, Kalita2023}, Rastall-Rainbow gravity \cite{Li_2024}, regularized 4D Einstein--Gauss--Bonnet gravity \cite{PRETEL2025}, and other extensions of general relativity (GR) \cite{Rajeev2026, Banerjee2017, Astashenok2023, Staicova2026}. These investigations have shown that the maximum mass of WDs is not universal but depends sensitively on the underlying theory of gravity. Although they demonstrate the potential of WDs as low-curvature laboratories for testing gravity, they also show that reproducing the observed mass--radius relation alone is insufficient to place quantitative constraints on a gravitational theory. Instead, robust constraints require a consistent treatment of the stellar microphysics, the gravitational mass definition, and the statistical comparison between theoretical predictions and observations.

Symmetric teleparallel gravity provides a geometric formulation of gravity in which the gravitational interaction is encoded in the non-metricity of the affine connection, with both curvature and torsion vanishing identically. In this gravitational formulation, the action linear in the non-metricity scalar $Q$ is dynamically equivalent to the Einstein--Hilbert action up to a boundary term, thereby providing the symmetric teleparallel counterpart of GR within the geometrical trinity of gravity \cite{BeltranJimenez2018Coincident,BeltranJimenez2019, Bahamonde_2023}. A natural extension is obtained by promoting the linear dependence on $Q$ to a general function $f(Q)$. This modification leads to gravitational dynamics that can depart from GR while preserving the geometric structure of symmetric teleparallel gravity \cite{Harko2018, BeltranJimenez2020Cosmology, Zhao2022, HEISENBERG20241}.

From a cosmological point of view, $f(Q)$ gravitational theory has attracted considerable interest as a geometrical alternative to the standard $\Lambda$CDM model for describing the late-time accelerated expansion of the universe. Observational studies have shown that viable $f(Q)$ models can provide satisfactory descriptions of the observed expansion history and competitive fits to cosmological data. In particular, redshift-based reconstructions have demonstrated the ability of $f(Q)$ models to account for the late-time acceleration \cite{Lazkoz2019}, while analyses of specific parameterizations using Markov Chain Monte Carlo methods have found observationally viable models with quintessence-like effective EoSs \cite{Koussour2023}. More recently, studies of hybrid $f(Q)$ models have shown that early-universe constraints can render their background expansion degenerate with $\Lambda$CDM \cite{Kolhatkar2026}. These results therefore support $f(Q)$ gravity as a viable geometrical extension of GR and motivate us to investigate its implications in other gravitational regimes, particularly in compact objects such as WDs.

A particularly simple function of this framework is the quadratic
extension \cite{Lu2019, LinZhai2021}
\begin{equation}\label{eq:modelintro}
f(Q)=Q+\alpha Q^2,
\end{equation}
which represents the lowest-order nonlinear correction to the linear
symmetric teleparallel equivalent of GR. The model introduces only one additional parameter $\alpha$ which controls the strength of the deviation from GR and has dimensions $[\alpha]=\mathrm{length}^2$. In the limit $\alpha\rightarrow0$, the standard GR dynamics are recovered, providing a well-defined reference case against which the effects of the nonlinear correction can be assessed. The quadratic form therefore offers a minimal extension of GR that allows the effects of modified gravity on the structure and properties of compact stars to be investigated with a single additional parameter, avoiding the larger parameter space associated with more general functional forms of $f(Q)$. Importantly, the model itself does not impose a preferred sign for $\alpha$. Since positive and negative values of the coupling modify the stellar structure equations in opposite directions \cite{LinZhai2021, FortesAraujo2026}, both branches should be considered in order to assess the full range of equilibrium configurations and to determine how the high-density stellar properties and the deviations from GR depend on the sign and magnitude of the free parameter $\alpha$.

It is worth noting that several previous studies of compact stars within $f(Q)$ gravity have adopted the linear form $f(Q)=aQ+b$ \cite{Gul2024, Maurya2024CQG, bhar2023dark, Ditta2024}. Nevertheless, because the linear $Q$ term is dynamically equivalent to the Einstein--Hilbert term up to a boundary contribution, the coefficient $a$ can be interpreted as a rescaling of the gravitational coupling, whereas $b$ plays the role of an effective cosmological contribution. Consequently, once the gravitational and cosmological constants are appropriately identified, the underlying dynamics of the linear model remain within the Einsteinian class. In contrast, the nonlinear form introduced in Eq.~\eqref{eq:modelintro} cannot be reduced to the Einstein--Hilbert form through such redefinitions, introducing additional contributions to the gravitational field equations and a nontrivial dependence on the non-metricity scalar. The quadratic model thus provides the simplest extension of the linear theory, with $\alpha$ controlling the leading correction to GR.

Spherically symmetric compact stars in $Q$-squared gravity have recently been studied in fixed metric ansatz constructions and covariant formulations \cite{LinZhai2021, Alwan2024, FortesAraujo2026, Heisenberg2026}. The WD problem has also recently been formulated in the quadratic $f(Q)$ model \cite{Sahoo2026}. These WD mass-radius calculations demonstrate that $\alpha$ modifies the equilibrium sequences relative to usual GR. To consistently assess the observational implications of both signs of the coupling, however, the positive and negative domains must be treated using the same prescriptions for the geometric setup, the matching to the vacuum exterior, and the statistical analysis. We therefore incorporate the affine equation into the stellar system, extend each accepted interior solution through its corresponding vacuum exterior, and confront the resulting configurations with observations using the ADM mass defined at spatial infinity. The integral of material energy and the asymptotic gravitational mass are calculated separately because they are distinct definitions in a nonlinear gravitational theory. The role of the mass definition has also been examined in related teleparallel compact star calculations \cite{FortesAraujo2023fT}.

The statistical problem cannot be reduced to selecting the closest mass and radius curve. The central density of each observed WD is unmeasured and is marginalized along the admitted sequence at fixed \(\alpha\). A logarithmic magnitude prior cannot pass continuously through \(\alpha=0\). We therefore define separate continuous models for \(\alpha<0\) and \(\alpha>0\), with GR as a discrete third model. Conditional posteriors determine the scale of \(|\alpha|\) within each sign, while the integrated evidences compare both signs with GR under explicit prior model weights. 

This work performs the signed inference on the regular branch of the general covariant spherical system. The branch follows from the affine equation and the regular center expansion. Each interior is matched to its exterior, and \(\MADM\) and \(\Rareal\) enter the likelihood. The observed masses and radii of Sirius B, QS Vir, V471 Tau B, ZTF J1901+1458, and LHS 4033 provide observational benchmarks for testing our theoretical mass--radius results, and these data are incorporated into our statistical analysis to constrain the parameter $\alpha$. Repeating the calculation with the first three objects measures the influence of the ultramassive stars on the result. Deterministic quadrature is used for the single continuous parameter in each sign sector and is tested against refinements of the central density and coupling grids. 

The paper is organized as follows. Sec.~\ref{sec:fq} gives the geometric conventions, spherical reduction, regular branch, mass definitions for our WD configurations, and validation tests. Sec.~\ref{sec:eos} defines the matter model and its reaction boundaries. Sec.~\ref{sec:likelihood} describes the observational sample and Bayesian likelihood. Sec.~\ref{sec:results} presents the resulting stellar sequences and the Bayesian analysis used to confront them with the WD observations. Finally, our conclusions are given in Sec.~\ref{sec:conclusions}.

\section{Quadratic \(f(Q)\) stellar structure}
\label{sec:fq}

\subsection{Geometric definitions and field equations}

We use the metric signature \((-+++)\) and the time coordinate \(x^0=ct\). The metric \(g_{\mu\nu}\) and affine connection \(\Gamma^\alpha{}_{\mu\nu}\) are independent variables. Symmetric teleparallel geometry imposes
\begin{equation}
R^\alpha{}_{\beta\mu\nu}(\Gamma)=0,
\qquad
T^\alpha{}_{\mu\nu}(\Gamma)=0,
\label{eq:flat-torsionless}
\end{equation}
while allowing the non-metricity tensor
\begin{equation}
Q_{\alpha\mu\nu}\equiv\nabla_\alpha g_{\mu\nu}
\label{eq:nonmetricity}
\end{equation}
to be nonzero. Its two traces are \(Q_\alpha=Q_{\alpha\ \mu}^{\ \mu}\) and \(\widetilde Q_\alpha=Q^\mu{}_{\alpha\mu}\). In the convention used here, the non-metricity conjugate is
\begin{equation}
\begin{split}
P^\alpha{}_{\mu\nu}
={}&-\frac14 Q^\alpha{}_{\mu\nu}
+\frac12 Q_{(\mu}{}^\alpha{}_{\nu)}
+\frac14\left(Q^\alpha-\widetilde Q^\alpha\right)g_{\mu\nu}\\
&-\frac14\delta^\alpha{}_{(\mu}Q_{\nu)},
\end{split}
\label{eq:superpotential}
\end{equation}
and the non-metricity scalar is given by
\begin{equation}
Q=Q_{\alpha\mu\nu}P^{\alpha\mu\nu}.
\label{eq:q-definition}
\end{equation}
These sign conventions fix the form of all subsequent reduced equations.

In $f(Q)$ gravity, the action is
\begin{equation}
S=\int d^4x\,\sqrt{-g}
\left[
\frac{f(Q)}{2\kappa}+\mathcal L_m
\right],
\qquad
\kappa=\frac{8\pi G}{c^4},
\label{eq:action}
\end{equation}
with minimally coupled matter. Metric variation can be written in terms of the Einstein tensor of the Levi--Civita connection \cite{Zhao2022}
\begin{equation}
f_Q\,\GEin_{\mu\nu}
+\frac12g_{\mu\nu}(Qf_Q-f)
+2f_{QQ}P^\alpha{}_{\mu\nu}\lc_\alpha Q
=\kappa T_{\mu\nu},
\label{eq:metric-field-equation}
\end{equation}
where \(f_Q=\partial f/\partial Q\), \(f_{QQ}=\partial^2f/\partial Q^2\), and the overcircle indicates that the corresponding quantity is constructed from the Levi--Civita connection. Independent variation of the flat connection with zero torsion yields
\begin{equation}
\nabla_\mu\nabla_\nu
\left(
\sqrt{-g}\,f_Q P_\alpha{}^{\mu\nu}
\right)=0.
\label{eq:affine-field-equation}
\end{equation}

For the functional form (\ref{eq:modelintro}), one gets
\begin{equation}
f_Q=1+2\alpha Q,
\qquad
f_{QQ}=2\alpha,
\qquad
Qf_Q-f=\alpha Q^2,
\label{eq:quadratic-derivatives}
\end{equation}
and, when $\alpha=0$, equations reduce identically to the Einstein equations for the conventions above.

\subsection{General spherical configuration}

The static spherical line element is written without fixing the areal radius to the affine radial coordinate, namely
\begin{equation}
ds^2=-e^{A(r)}(dx^0)^2+e^{B(r)}dr^2+C(r)^2d\Omega^2.
\label{eq:general-metric}
\end{equation}
Thus, \(C(r)\) is the areal radius function: a two sphere at coordinate \(r\) has area \(4\pi C^2\). The flat inertial connection with zero torsion is the Levi--Civita connection of Minkowski spacetime expressed in spherical coordinates. Its nonzero independent components are
\begin{align}
\Gamma^{r}{}_{\theta\theta}&=-r,
&
\Gamma^{r}{}_{\phi\phi}&=-r\sin^2\theta,
\nonumber\\
\Gamma^{\theta}{}_{r\theta}
=\Gamma^{\theta}{}_{\theta r}&=\frac1r,
&
\Gamma^{\theta}{}_{\phi\phi}&=-\sin\theta\cos\theta,
\nonumber\\
\Gamma^{\phi}{}_{r\phi}
=\Gamma^{\phi}{}_{\phi r}&=\frac1r,
&
\Gamma^{\phi}{}_{\theta\phi}
=\Gamma^{\phi}{}_{\phi\theta}&=\cot\theta .
\label{eq:inertial-connection}
\end{align}
This connection is nonzero in spherical coordinates but becomes zero under the corresponding Cartesian coordinate transformation. Retaining \(C(r)\) distinguishes the general metric ansatz from the special choice \(C=r\). In nonlinear \(f(Q)\), the latter cannot be imposed before the affine equation is checked \cite{Zhao2022}.

Substitution of Eqs.~(\ref{eq:general-metric}) and (\ref{eq:inertial-connection}) into Eq.~(\ref{eq:q-definition}) gives the non-metricity scalar
\begin{equation}
\begin{split}
Q=-\frac{e^{-B}}{r^2C^2}
\big[&
r^3e^B(A'+B')
-2r^2CA'C'
-2r^2C'^2\\
&+rC^2(A'-B')
+4rCC'
-2C^2
\big].
\end{split}
\label{eq:q-general}
\end{equation}
The metric equations are evaluated directly from Eq.~(\ref{eq:metric-field-equation}). For traceability, the geometric tensors entering them are
\begin{align}
\GEin_{tt}
&=
\frac{e^{A-B}}{C^2}
\left(CB'C'-2CC''+e^B-C'^2\right),
\label{eq:gtt-general}\\
\GEin_{rr}
&=
\frac{CA'C'-e^B+C'^2}{C^2},
\label{eq:grr-general}\\
\GEin_{\theta\theta}
&=
\frac{Ce^{-B}}4
\left[
C(A'^2-A'B'+2A'')
\right.
\nonumber\\[-2pt]
&\hspace{3.5em}\left.
+2A'C'-2B'C'+4C''
\right],
\label{eq:gtheta-general}
\end{align}
and the required radial components of \(P^\alpha{}_{\mu\nu}\) are
\begin{align}
P^r{}_{tt}
&=
\frac{e^{A-B}}{2rC^2}
\left(r^2e^B-2rCC'+C^2\right),
\label{eq:prtt}\\
P^r{}_{rr}
&=
-\frac{r^2e^B-C^2}{2rC^2},
\label{eq:prrr}\\
P^r{}_{\theta\theta}
&=
\frac{Ce^{-B}}{4r}
\left(rCA'+2rC'-2C\right).
\label{eq:prtheta}
\end{align}
Equations~(\ref{eq:q-general}) through (\ref{eq:prtheta}), together with Eq.~(\ref{eq:metric-field-equation}), specify all three diagonal metric equations without an implicit gauge replacement.

We model the matter-energy source as a perfect fluid, whose energy-momentum tensor is given by
\begin{equation}
T^\mu{}_\nu=\mathrm{diag}(-\Ecal,p,p,p),
\label{eq:perfect-fluid}
\end{equation}
where \(\Ecal\) is the total energy density and \(p\) is the isotropic pressure of the fluid. For minimally coupled matter, the energy-momentum tensor is covariantly conserved, i.e., \(\lc_\mu T^\mu{}_\nu=0\). The radial component of this conservation equation yields the hydrostatic equilibrium condition
\begin{equation}
\frac{dp}{dr}=-\frac{\Ecal+p}{2}A'.
\label{eq:hydrostatic}
\end{equation}
The EoS supplies both \(p\) and \(\Ecal\), so the gravitational source is determined by the full energy density rather than by the rest-mass density alone.

\subsection{Affine equation and regular branch}

For the spherical ansatz, Eq.~(\ref{eq:affine-field-equation}) reduces to \cite{Zhao2022}
\begin{equation}
\begin{split}
\big[&
(e^B r^2-C^2)(4+rA'+rB')\\
&+4C^2-4rCC'+2C^2rB'
\big]f_Q'\\
&+2(e^B r^2-C^2)r f_Q''=0.
\end{split}
\label{eq:affine-spherical}
\end{equation}
Defining
\begin{equation}
D=e^B r^2-C^2,
\label{eq:D}
\end{equation}
and using \(f_Q'=2\alpha Q'\) for \(\alpha\neq0\), Eq.~(\ref{eq:affine-spherical}) can be rearranged exactly as
\begin{equation}
2(DQ')'+(A'-B')DQ'=0.
\label{eq:affine-total-derivative}
\end{equation}
Its first integral reads
\begin{equation}
DQ'=K e^{(B-A)/2},
\label{eq:affine-first-integral}
\end{equation}
where \(K\) is a constant.

The physical branch is fixed by regularity conditions at the center, rather than by fitting a prescribed stellar profile. Near the center, we write
\begin{align}
A&=A_0+a_2r^2+O(r^4),
&
B&=b_2r^2+O(r^4),
\nonumber\\
C&=r+c_3r^3+O(r^5),
&
Q&=q_2r^2+O(r^4).
\label{eq:center-series}
\end{align}
The leading metric equations give
\begin{align}
a_2&=\frac{\kappa}{2}
\left(p_c+\frac{\Ecal_c}{3}\right),
\label{eq:a2}\\
b_2&=\frac{\kappa\Ecal_c}{3}+6c_3,
\label{eq:b2}\\
q_2&=-\frac13
\left(
120c_3^2
+20\kappa\Ecal_c c_3
+\kappa^2\Ecal_c^2
+\kappa^2\Ecal_c p_c
\right),
\label{eq:q2}
\end{align}
where $\Ecal_c$ and $p_c$ denote the energy density and pressure at the center of the star, respectively. The leading affine coefficient is
\begin{equation}
10q_2(b_2-2c_3)=0.
\label{eq:center-affine}
\end{equation}

For \(\Ecal_c>0\) and \(p_c>0\), the quadratic equation obtained from \(q_2=0\) has no real \(c_3\). The real analytic branch therefore satisfies \(b_2=2c_3\), which gives
\begin{equation}
c_3=-\frac{\kappa\Ecal_c}{12},
\qquad
b_2=-\frac{\kappa\Ecal_c}{6}.
\label{eq:center-branch-coefficients}
\end{equation}
Regularity also sets \(K=0\) in Eq.~(\ref{eq:affine-first-integral}). Since the constant-\(Q\) alternative is excluded by the center series, the stellar branch obeys \(D=0\), and hence
\begin{equation}
C(r)=r e^{B(r)/2}.
\label{eq:regular-affine-branch}
\end{equation}
Equation~(\ref{eq:regular-affine-branch}) is consequently a result of the field equations and boundary conditions, rather than being introduced as an additional phenomenological ansatz.

\subsection{Reduced stellar system}

On the branch (\ref{eq:regular-affine-branch}), it is convenient to define
\begin{equation}
U=A',
\qquad
W=B',
\qquad
H=U+W, 
\label{eq:reduced-variables}
\end{equation}
so that the areal derivative and non-metricity scalar become
\begin{align}
C'&=e^{B/2}\left(1+\frac{rW}{2}\right),
\label{eq:cprime}\\
Q&=\frac12e^{-B}W(2U+W)
=e^{-B}W\left(H-\frac{W}{2}\right),
\label{eq:q-reduced}\\
Q'&=e^{-B}
\left[
WU'+(U+W)W'
\right]-QW.
\label{eq:qprime-reduced}
\end{align}
The three diagonal field equations reduce to
\begin{align}
&-f_Qe^{-B}
\left(\frac{W^2}{4}+W'+\frac{2W}{r}\right)
\nonumber\\[-2pt]
&\qquad
-\frac{\alpha Q^2}{2}
-2\alpha e^{-B}WQ'
=\kappa\Ecal,
\label{eq:ett-reduced}\\
&\frac{f_Qe^{-B}}{4r}
\left(2rUW+rW^2+4U+4W\right)
\nonumber\\[-2pt]
&\qquad
+\frac{\alpha Q^2}{2}
=\kappa p,
\label{eq:err-reduced}\\
&\frac{f_Qe^{-B}}4
\left[
U^2+2U'+2W'+\frac{2(U+W)}{r}
\right]
\nonumber\\[-2pt]
&\qquad
+\frac{\alpha Q^2}{2}
+\alpha e^{-B}(U+W)Q'
=\kappa p.
\label{eq:etheta-reduced}
\end{align}
The radial equation (\ref{eq:err-reduced}) is algebraic in \(H\) after Eq.~(\ref{eq:q-reduced}) is inserted. For \(\alpha\neq0\), the resulting quadratic constraint has two algebraic roots for \(H\). We retain the root that remains finite as \(\alpha\to0\) and reduces to the solution of the radial Einstein equation. This limiting condition selects the root with a regular GR limit, although the solutions at finite \(\alpha\) satisfy the full quadratic \(f(Q)\) equations and generally differ from their GR counterparts. Differentiating this constraint and combining it with Eq.~(\ref{eq:ett-reduced}) determines \(H'\) and \(W'\); then \(U'=H'-W'\). Equation~(\ref{eq:etheta-reduced}) is not discarded. It is evaluated independently as a closure identity. The conservation equation \eqref{eq:hydrostatic} can be evolved through a monotonic EoS coordinate \(\rho\), namely
\begin{equation}
\rho'=
-\frac{(\Ecal+p)U}{2\,dp/d\rho}.
\label{eq:rho-evolution}
\end{equation}
This formulation enforces the radial constraint locally and does not require a Newton Raphson iteration at each radius.

At the center, \(A_0\) is an arbitrary time normalization, \(B(0)=0\), \(U(0)=W(0)=0\), and \(\rho(0)=\rho_c\). A fourth order regular expansion starts the integration at a finite radius. The stellar surface \(r_s\) is the first point at which the cold EoS reaches \(p=0\), and the physical radius is
\begin{equation}
\Rareal=C(r_s).
\label{eq:areal-radius}
\end{equation}
Only intervals that remain regular, preserve \(C'>0\), and connect to a convergent asymptotically flat exterior are admitted. Since asymptotic flatness implies \(Q\to0\) and hence \(f_Q\to1\), we also require that \(f_Q\) does not vanish along the solution. This condition keeps the stellar configuration in the nondegenerate sector continuously connected to the asymptotic GR limit and, by continuity, implies \(f_Q>0\) throughout the admitted domain \cite{BeltranJimenez2020Cosmology}.

\subsection{Exterior matching and mass definitions}

The equivalent of integrated energy-density contribution is defined using the areal-coordinate measure,
\begin{equation}
\frac{d\Mint}{dr}
=4\pi\frac{\Ecal}{c^2}C^2C',
\qquad
\Mint(0)=0.
\label{eq:material-mass}
\end{equation}
Evaluated at the stellar surface, \(\Mint(r_s)\) is the total matter-energy density integrated over the areal-coordinate measure and expressed in mass units. It quantifies the integrated material content and is retained only as a diagnostic. It differs from the baryonic mass, which is constructed from the conserved baryon number using the proper spatial volume, and from the ADM mass, which is determined by the asymptotic geometry.

The geometric mass function required to obtain the latter is defined directly from the areal radius \(C\) by
\begin{equation}
1-\frac{2G\Mgeom}{c^2C}
=g^{ab}(\partial_aC)(\partial_bC)
=e^{-B}C'^2,
\label{eq:geometric-mass-definition}
\end{equation}
where \(a,b\in\{x^0,r\}\). On the regular branch,
\begin{equation}
\Mgeom(r)
=\frac{c^2C}{2G}
\left[
1-\left(1+\frac{rB'}2\right)^2
\right].
\label{eq:geometric-mass}
\end{equation}
The stellar interior is continued through the vacuum equations in the same affine branch. The additive constant in \(A\) is chosen such that \(A\to0\) asymptotically, and a constant asymptotic factor in \(C/r\) is removed by radial normalization. The observable mass is
\begin{equation}
\MADM=\lim_{C\rightarrow\infty}\Mgeom(C),
\label{eq:adm-mass}
\end{equation}
which corresponds to the total gravitational mass of the asymptotically flat spacetime, in analogy with the gravitational mass extracted from asymptotic exterior solutions in other modified gravity theories \cite{ASTASHENOK2015160, Astashenok2017, Pretel2022}. In the nonlinear $f(Q)$ model, $\Mgeom(r_s)$ need not equal its asymptotic limit because the vacuum metric can carry a finite effective modified gravity contribution before approaching its asymptotic value. The production criterion is convergence of the asymptotic mass, not pointwise constancy from the material surface. Accordingly, the theoretical sequence used in the likelihood is
\begin{equation}
\mathcal S_\alpha
=
\left\{
\MADM(\rho_c;\alpha), \Rareal(\rho_c;\alpha)
\right\}.
\label{eq:observable-sequence}
\end{equation}

\subsection{Numerical construction and validation}
\label{sec:numerical-validation}

The signed production set contains 37 negative-\(\alpha\) sequences, the GR sequence, and 37 positive-\(\alpha\) sequences, for a total of 75 coupling values spanning \(10^{14}\leq|\alpha|/\mathrm{cm^2}\leq10^{21}\). The combined library contains 4529 admitted stellar configurations, with at least 59 directly integrated central-density nodes in each sequence. The interior equations are integrated adaptively with relative tolerance \(5\times10^{-7}\) and a maximum radial step of \(15\,\mathrm{km}\). The positive-\(\alpha\) exteriors are continued to \(30r_s\), while the negative-\(\alpha\) exteriors are continued to \(50r_s\), both with relative tolerance \(5\times10^{-8}\). In both sectors, admission requires the relative variation of the geometric mass in the exterior tail to remain below \(10^{-6}\). Eighteen negative-\(\alpha\) configurations spanning the coupling and central-density domains were repeated with relative tolerance \(10^{-8}\), a maximum interior step of \(2\,\mathrm{km}\), an exterior extending to \(50r_s\), and exterior relative tolerance \(10^{-10}\).

The symbolic reduction reproduces the \(C=r\) equations of Ref.~\cite{Zhao2022}, the corresponding affine equation, and the Einstein tensor at \(\alpha=0\). An independent GR TOV calculation agrees with the present GR mass and radius to \(5.96\times10^{-6}\) and \(6.96\times10^{-6}\), respectively. Across the validation cases, tightening the integration changes mass by at most \(1.12\times10^{-5}\) and radius by at most \(8.79\times10^{-6}\). The normalized radial constraint, angular identity, and evolution closure residuals remain below \(2.5\times10^{-16}\) in the audited core profiles. The maximum adaptive interpolation error in the final sequence set is \(8.57\times10^{-4}\).

For the primary analysis, each sequence is restricted to the part preceding its first mass maximum. This is a conservative branch mask motivated by the turning structure of the equilibrium family. It is not a radial mode calculation and is not called a proof of dynamical stability.

\section{Equation of state}
\label{sec:eos}

\subsection{Cold degenerate matter}

The primary matter model assumes cold, nonrotating, nonmagnetic \(\Ctwelve\) matter. The notation identifies the adopted composition; it does not assert that every star in the observational sample is composed purely of $\Ctwelve$ matter. Charge neutrality gives
\begin{equation}
n_e=Y_en_b,
\qquad
Y_e=\frac ZA=\frac12.
\label{eq:charge-neutrality}
\end{equation}
The electron Fermi momentum is represented by
\begin{equation}
x=\frac{p_F}{m_ec}
=\lambda_e(3\pi^2n_e)^{1/3},
\qquad
\lambda_e=\frac{\hbar}{m_ec},
\label{eq:x}
\end{equation}
and the electron chemical potential, including rest energy, is
\begin{equation}
\mu_e=m_ec^2\sqrt{1+x^2}.
\label{eq:electron-chemical-potential}
\end{equation}

The ideal degenerate electron pressure and total electron energy density are
\begin{align}
p_e&=
\frac{m_ec^2}{8\pi^2\lambda_e^3}
\left[
x\left(\frac{2x^2}{3}-1\right)\sqrt{1+x^2}
+\operatorname{asinh}x
\right],
\label{eq:electron-pressure}\\
\Ecal_e&=
\frac{m_ec^2}{8\pi^2\lambda_e^3}
\left[
x(1+2x^2)\sqrt{1+x^2}
-\operatorname{asinh}x
\right].
\label{eq:electron-energy}
\end{align}
The Coulomb lattice contribution is written
\begin{equation}
\Ecal_L=C_L e^2 n_e^{4/3}f(Z),
\qquad
p_L=\frac{\Ecal_L}{3},
\label{eq:lattice}
\end{equation}
where the adopted body centered cubic convention has \(C_L<0\) and \(f(Z)=Z^{2/3}\) for the one component layer \cite{Salpeter1961,HamadaSalpeter1961}. The thermodynamic quantities entering the stellar equations are
\begin{equation}
p=p_e+p_L,
\qquad
\Ecal=\Ecal_{\rm rest}+\Ecal_e+\Ecal_L.
\label{eq:total-eos}
\end{equation}
Here \(\Ecal_{\rm rest}\) is the adopted ionic or atomic rest energy contribution. Equation~(\ref{eq:total-eos}), rather than rest mass density alone, supplies the source in Eqs.~(\ref{eq:ett-reduced}) and (\ref{eq:hydrostatic}).

\subsection{Inverse beta instability and neutron drip}

The high density boundary is set by the composition dependent weak interaction threshold. The Gibbs free energy per baryon is denoted by
\begin{equation}
\Gcal=\frac{\Ecal+p}{n_b}.
\label{eq:gibbs}
\end{equation}
For an electron capture channel
\begin{equation}
(A,Z)+\Delta Z\,e^-
\rightarrow
(A,Z-\Delta Z)+\Delta Z\,\nu_e,
\label{eq:capture}
\end{equation}
the atomic mass threshold is
\begin{equation}
\mu_e^\beta=
\frac{
\left[
M_{\rm atom}(A,Z-\Delta Z)-M_{\rm atom}(A,Z)
\right]c^2
}{\Delta Z}
+m_ec^2.
\label{eq:capture-threshold}
\end{equation}
The transition is located by equality of the parent and daughter \(\Gcal\) at fixed pressure, including the lattice correction \cite{Chamel2013}. For the adopted \(\Ctwelve\) model this boundary occurs near
\begin{equation}
\rho_\beta\simeq4.16\times10^{10}\,
\mathrm{g\,cm^{-3}},
\qquad
\mu_e\simeq14.18\,\mathrm{MeV}.
\label{eq:beta-numbers}
\end{equation}
These values delimit the EoS domain; they do not constitute a dynamical collapse calculation.

Neutron drip is a distinct transition at which the equilibrium matter begins to contain free neutrons \cite{Baym1971}. It lies beyond the regime without free neutrons used here. The primary sequences are therefore stopped before a neutron rich matter model would be required. Related WD calculations show that electron capture and pycnonuclear thresholds can strongly restrict massive sequences when composition, magnetic fields, or rotation are varied \cite{Otoniel2019Magnetized,Malheiro2021Nuclear}. Temperature, rotation, magnetic pressure, composition gradients, and alternative compositions are not marginalized in the present inference.

\section{Observational sample and likelihood}
\label{sec:likelihood}

\subsection{Reference and sensitivity samples}

The reference sample contains Sirius B, QS Vir, V471 Tau B, ZTF J1901+1458, and LHS 4033. The first three objects have the most direct mass determinations in the adopted set. Sirius B combines an astrometric dynamical mass with a photometric and spectroscopic radius \cite{Bond2017,JoyceSirius2018}. QS Vir has mass and radius constraints from an eclipsing binary analysis \cite{Parsons2016}, and V471 Tau B is included with its revised binary parameters \cite{Muirhead2022}.

The two ultramassive objects extend the reference sample into the compact region where the signed sequences have their largest separation. ZTF J1901+1458 is rapidly rotating and strongly magnetized, and its reported mass spans a composition dependent interval \cite{Caiazzo2021}. LHS 4033 has a parallax based radius and a mass interval obtained with theoretical mass--radius relations \cite{Dahn2004}. Their inclusion supplies sensitivity to the high density branch sampled by the signed model. We repeat the inference with Sirius B, QS Vir, and V471 Tau B alone to measure how the posterior changes when the ultramassive objects are removed. This three object calculation is the direct sample sensitivity test.

\begin{table*}[t]
\caption{Five object reference sample. Symmetric uncertainties are shown with \(\pm\); asymmetric uncertainties are shown as lower and upper values. The ZTF J1901+1458 and LHS 4033 masses enter as uniform published ranges. The subset column identifies the three objects used in the direct sample sensitivity test and the two ultramassive objects that extend the reference analysis.}
\label{tab:sample}
\begin{ruledtabular}
\begin{tabular}{lcccccc}
Object
& \(M/\Msun\)
& mass model
& \(R\) (km)
& radius uncertainty (km)
& subset
& reference\\
\colrule
Sirius B
& \(1.018\pm0.011\)
& split normal
& \(5633.8\)
& \(\pm32.0\)
& direct
& \cite{Bond2017,JoyceSirius2018}\\
QS Vir
& \(0.782\pm0.013\)
& split normal
& \(7430.1\)
& \(\pm48.7\)
& direct
& \cite{Parsons2016}\\
V471 Tau B
& \(0.797\pm0.016\)
& split normal
& \(7931.0\)
& \(\pm410.5\)
& direct
& \cite{Muirhead2022}\\
ZTF J1901+1458
& \(1.327\)--\(1.365\)
& uniform range
& \(2140\)
& \(-230,+160\)
& ultramassive
& \cite{Caiazzo2021}\\
LHS 4033
& \(1.310\)--\(1.335\)
& uniform range
& \(2560.2\)
& \(\pm90.4\)
& ultramassive
& \cite{Dahn2004}\\
\end{tabular}
\end{ruledtabular}
\end{table*}

\subsection{Sequence likelihood}

For a measured quantity \(x_i\) with lower and upper uncertainties \(\sigma_{i,-}\) and \(\sigma_{i,+}\), we use the normalized split-normal density
\begin{equation}
\ln\mathcal N_{\rm split}(x|x_i)
=
\ln\left[
\frac{\sqrt{2/\pi}}{\sigma_{i,-}+\sigma_{i,+}}
\right]
-\frac{(x-x_i)^2}{2\sigma_{i,s}^2},
\label{eq:split-normal}
\end{equation}
where \(\sigma_{i,s}=\sigma_{i,-}\) for \(x<x_i\) and \(\sigma_{i,+}\) otherwise. The mass and radius factors are taken to be independent because numerical covariance matrices are not available for the adopted measurements. This zero-covariance choice is an explicit likelihood approximation.

For Sirius B, QS Vir, and V471 Tau B,
\begin{equation}
\ln\mathcal L_i(M,R)
=
\ln\mathcal N_{\rm split}(M|M_i)
+\ln\mathcal N_{\rm split}(R|R_i).
\label{eq:primary-likelihood}
\end{equation}
For ZTF J1901+1458 and LHS 4033, the radius retains Eq.~(\ref{eq:split-normal}), while the mass likelihood is uniform and normalized over its published interval and zero outside.

At fixed \(\alpha\), the nuisance central density is marginalized over the admitted domain \(\mathcal D_\alpha\):
\begin{equation}
\begin{split}
\mathcal L_i(\alpha)
={}&
\int_{\mathcal D_\alpha}
\mathcal L_i
\left[
\MADM(\rho_c;\alpha),
\Rareal(\rho_c;\alpha)
\right]\\
&\times\pi(\ln\rho_c|\alpha)\,d\ln\rho_c,
\end{split}
\label{eq:rho-marginalization}
\end{equation}
where the prior \(\pi(\ln\rho_c|\alpha)\) is uniform and normalized on the selected branch. For a stellar sample \(\mathcal S\), the joint likelihood is
\begin{equation}
\ln\mathcal L_{\mathcal S}(\alpha)
=\sum_{i\in\mathcal S}\ln\mathcal L_i(\alpha).
\label{eq:joint-likelihood}
\end{equation}
The reference likelihood, denoted by \(\mathcal L_{\mathcal S_5}\), contains all five objects listed in Table~\ref{tab:sample}. The sensitivity likelihood \(\mathcal L_{\mathcal S_3}\) contains Sirius B, QS Vir, and V471 Tau B. Mass and radius along each directly integrated sequence are interpolated with a monotonic piecewise cubic Hermite interpolant in \(\ln\rho_c\). Each nuisance density integral uses 4097 quadrature points. Failed intervals, points outside the admitted domain, and disconnected branches are never bridged.

\subsection{Signed prior and model comparison}

The signed analysis contains three mutually exclusive models,
\begin{equation}
\mathcal M_-:\alpha<0,
\qquad
\mathcal M_{\rm GR}:\alpha=0,
\qquad
\mathcal M_+:\alpha>0.
\label{eq:signed-models}
\end{equation}
The continuous sectors share the magnitude coordinate
\begin{equation}
x\equiv\log_{10}\!\left(\frac{|\alpha|}{\mathrm{cm^2}}\right),
\qquad 14\leq x\leq21,
\label{eq:signed-alpha-coordinate}
\end{equation}
with conditional priors
\begin{equation}
\pi(x|\mathcal M_-)=\pi(x|\mathcal M_+)=\frac17.
\label{eq:alpha-prior}
\end{equation}
The signed mapping is
\begin{equation}
\alpha_s(x)=
\begin{cases}
-10^x\,\mathrm{cm^2}, & s=-,\\
+10^x\,\mathrm{cm^2}, & s=+.
\end{cases}
\label{eq:signed-alpha-mapping}
\end{equation}
GR is a discrete model and is not represented by a bin at the origin of either logarithmic prior.

For \(s\in\{-,+\}\), the posterior conditional on a sign sector is
\begin{equation}
p(x|d,\mathcal M_s)
=
\frac{\mathcal L_{\mathcal S}[\alpha_s(x)]
\pi(x|\mathcal M_s)}{\mathcal Z_s},
\label{eq:signed-conditional-posterior}
\end{equation}
where
\begin{align}
\mathcal Z_s
&=\int_{14}^{21}
\mathcal L_{\mathcal S}[\alpha_s(x)]
\pi(x|\mathcal M_s)\,dx,
\nonumber\\
\mathcal Z_{\rm GR}
&=\mathcal L_{\mathcal S}(0)
\label{eq:signed-evidences}
\end{align}
are the sector evidences. Equal prior model weights are assigned in the reference comparison,
\begin{equation}
P(\mathcal M_-)=P(\mathcal M_{\rm GR})=P(\mathcal M_+)=\frac13.
\label{eq:sector-prior-weights}
\end{equation}
The posterior probability of each sector is then
\begin{equation}
P(\mathcal M_k|d)
=
\frac{\mathcal Z_k P(\mathcal M_k)}
{\sum_{j\in\{-,{\rm GR},+\}}\mathcal Z_jP(\mathcal M_j)},
\label{eq:sector-posterior-probability}
\end{equation}
and the pairwise log Bayes factor is
\begin{equation}
\ln B_{i/j}=\ln\mathcal Z_i-\ln\mathcal Z_j.
\label{eq:bayes-factor}
\end{equation}
The conditional posterior in Eq.~(\ref{eq:signed-conditional-posterior}) locates the coupling magnitude within a fixed sign sector. Equation~(\ref{eq:sector-posterior-probability}) compares the two signs with GR. Direct deterministic quadrature is used because each continuous model has one parameter. It also permits explicit tests of the density quadrature and coupling grid resolution \cite{Sivia2006}.

\section{Results and discussion}
\label{sec:results}

The signed extension contains 37 negative couplings, GR, and 37 positive couplings over \(10^{14}\leq |\alpha|/\mathrm{cm^2}\leq10^{21}\). Each configuration satisfies the regular center conditions, the algebraic radial constraint, \(f_Q>0\), monotonicity of the areal radius, and convergence of the exterior mass. These requirements define the admitted equilibrium domain of the selected affine branch. They do not establish radial stability.

\begin{figure*}[t]
\includegraphics[width=\textwidth]{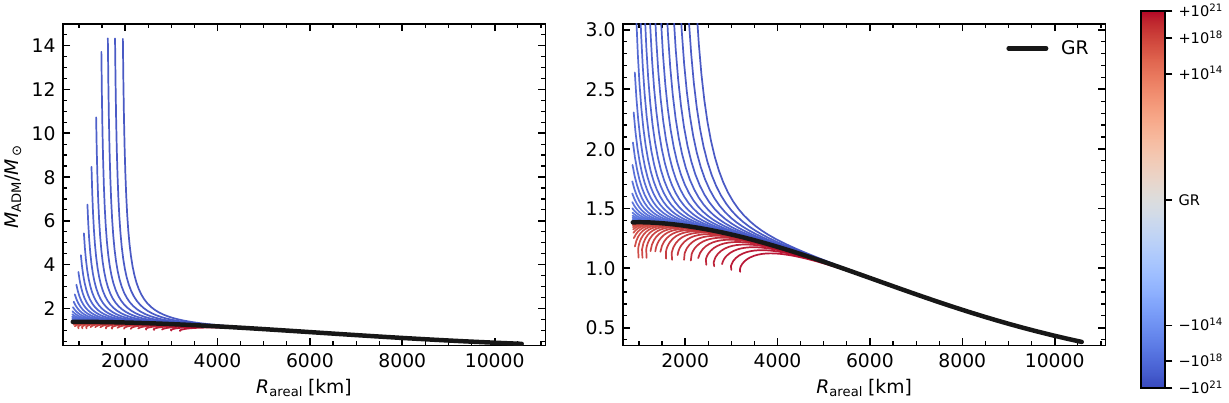}
\caption{\label{fig:mr-family}ADM mass and areal radius for the complete signed coupling family. The left panel displays every admitted equilibrium configuration, including the high mass extensions found for negative \(\alpha\). The right panel restricts the mass range while retaining the full radius interval. The color scale encodes the sign and magnitude of \(\alpha\), and the black curve denotes GR. Existence in the admitted numerical domain does not imply radial stability.}
\end{figure*}

\begin{figure}[b]
\includegraphics[width=\columnwidth]{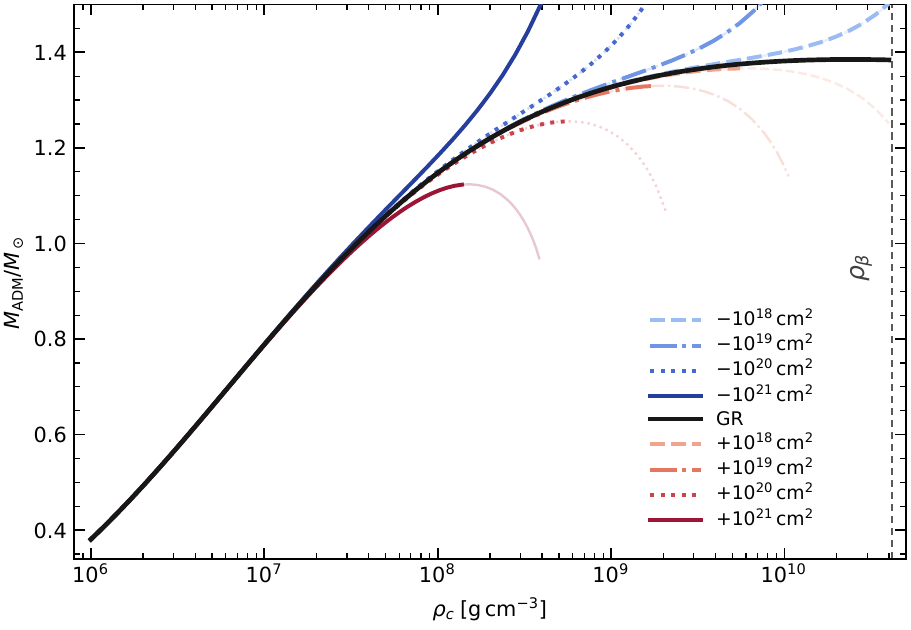}
\caption{\label{fig:mass-density}ADM mass as a function of central density for representative positive and negative couplings and for GR. The vertical line denotes the inverse beta threshold of the adopted cold carbon EoS. Curves that cross the upper plotting boundary continue within the admitted equilibrium domain shown in Fig.~\ref{fig:mr-family}. No marked point is interpreted as a stellar stability boundary via small radial perturbations.}
\end{figure}

Figure~\ref{fig:mr-family} shows how the sign of the quadratic term changes the compact part of the mass and radius relation. All sectors approach the same low density curve because the quadratic contribution is controlled by \(\alpha Q\) and becomes small when the non-metricity scale is low. The separation develops as the central density and compactness increase. Positive couplings shift the terminal configurations toward larger radii and lower masses, while the mass--radius sequence still develops a maximum-mass configuration characterized by $dM/d\rho_c=0$ (see also Fig.~\ref{fig:mass-density}). Negative couplings act in the opposite direction, modifying the Chandrasekhar mass limit and allowing the maximum mass to increase beyond its GR counterpart. In this case, however, the equilibrium sequence does not exhibit a finite turning point with $dM/d\rho_c=0$ for $\vert\alpha\vert$ large enough, and the mass continues to grow along the sequence. This behavior is also found in compact-star configurations made of polytropic fluid within $f(Q)$ gravity \cite{LinZhai2021}, particularly in studies of neutron stars with negative values of $\alpha$.

The negative extension is not a small perturbation throughout the explored domain. The largest mass in the calculated family is \(14.34\,\Msun\), obtained near \(\alpha=-4.64\times10^{20}\,\mathrm{cm^2}\), \(R_{\rm areal}=1626\,\mathrm{km}\), and \(\rho_c=3.15\times10^{10}\,\mathrm{g\,cm^{-3}}\). This value reports a regular equilibrium solution that passes the adopted interior and exterior checks. It must not be interpreted as a stable WD mass because no radial eigenmode calculation has been performed. The right panel of Fig.~\ref{fig:mr-family} restricts the vertical range to the usual WD window and shows that the same sign dependence is already present near the GR mass scale.

\begin{figure}[t]
\includegraphics[width=\columnwidth]{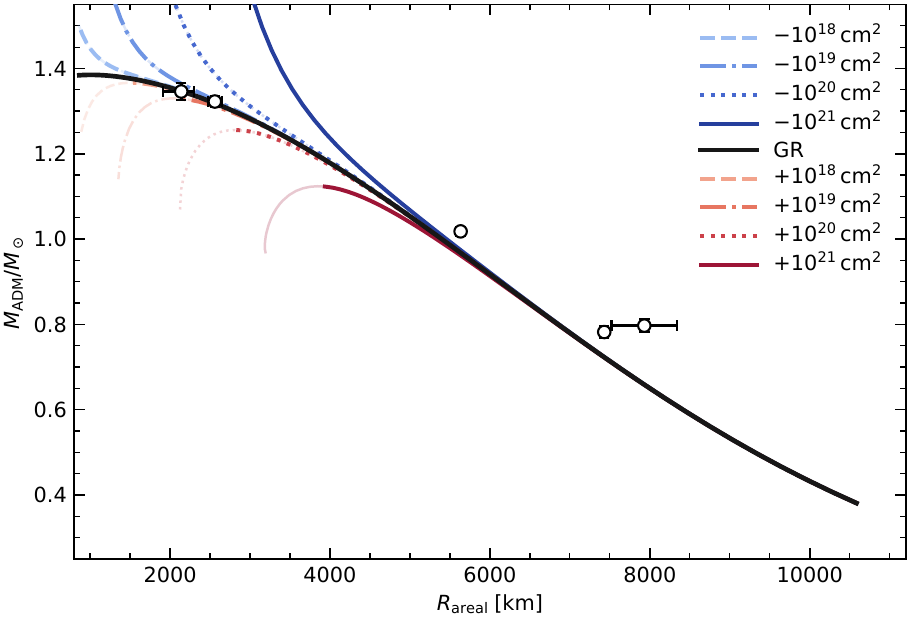}
\caption{\label{fig:mr-observations}Representative signed coupling sequences compared with the five mass--radius measurements. The displayed couplings are \(\alpha=\pm10^{18},\pm10^{19},\pm10^{20},\pm10^{21}\,\mathrm{cm^2}\), together with GR. Dark segments identify the conservative portion used in the reference likelihood, while faint continuations show the remaining admitted equilibrium branch. All observations are plotted with their mass and radius uncertainties. Parameter inference uses the sequence likelihood marginalized over central density rather than the distance to a visually selected curve.}
\end{figure}

Figure~\ref{fig:mr-observations} compares representative magnitudes of both signs with the five adopted objects. The values \(|\alpha|=10^{18},10^{19},10^{20},10^{21}\,\mathrm{cm^2}\) were selected to expose the sign dependence rather than to represent posterior summaries. Negative couplings raise the compact branch relative to GR. Positive couplings lower its turning scale and can terminate before reaching the largest central densities allowed by the matter model. The separation is strongest at radii below approximately \(4000\,\mathrm{km}\), where ZTF J1901+1458 and LHS 4033 are located. Sirius B, QS Vir, and V471 Tau B occupy a region in which most representative curves remain close to GR.

The five measurements therefore do not contribute equal leverage on \(\alpha\). The ultramassive objects probe the region of strongest theoretical separation, but their interpretation also carries the largest mismatch with the present stellar model. ZTF J1901+1458 is rapidly rotating and strongly magnetized, while the calculated configurations are static and nonmagnetic. The mass interval of LHS 4033 has a larger dependence on theoretical mass and radius relations than the three direct objects. We retain all five observed objects in the reference signed analysis and quantify the change relative to the three object sample below. Their placement in Fig.~\ref{fig:mr-observations} is in the sample comparison, not an independent validation of the gravity model.

The dependence on central density clarifies the origin of the two trends. Figure~\ref{fig:mass-density} shows that the sectors are nearly indistinguishable at low \(\rho_c\). As \(\rho_c\) increases, the negative sequences continue upward through the upper boundary of the displayed mass interval, whereas positive sequences bend toward lower masses or terminate at a branch condition. The vertical line marks the inverse beta threshold \(\rho_\beta\simeq4.16\times10^{10}\,\mathrm{g\,cm^{-3}}\). A sequence can reach this matter boundary only if the geometric and exterior conditions remain admissible up to that density. At large positive coupling, the radial constraint discriminant becomes restrictive before \(\rho_\beta\). For negative coupling, the configurations shown in Fig.~\ref{fig:mr-family} can remain admitted up to the high density part of the carbon EoS.

Maximum mass symbols are not included in Fig.~\ref{fig:mass-density}. A turning point along a one-parameter equilibrium family is useful for defining the conservative integration mask, but it is not equivalent to a radial-mode stability analysis in the present modified gravity system. The figure is instead used to identify which density interval generates the observable separation and which physical or geometric condition terminates each sequence.

The exterior calculation fixes which mass enters the observational comparison. Across the conservative signed sequences, the integrated material mass and the ADM mass remain numerically close. For the positive production family the maximum relative difference satisfies
\begin{equation}
\delta_M^{\max}
\equiv
\max_{\rho_c}
\frac{|M_{\rm int}-M_{\rm ADM}|}{M_{\rm ADM}}
\leq1.44\times10^{-5}.
\label{eq:mass-audit-result}
\end{equation}
Replacing \(M_{\rm ADM}\) by \(M_{\rm int}\) in the signed three object analysis changes the sector probabilities by less than \(2\times10^{-5}\). This agreement is a numerical consistency result over the admitted domain, not an exact identity between the two mass definitions. We retain \(M_{\rm ADM}\) because it is determined by the asymptotic exterior geometry.

\begin{figure*}[t]
\includegraphics[width=\textwidth]{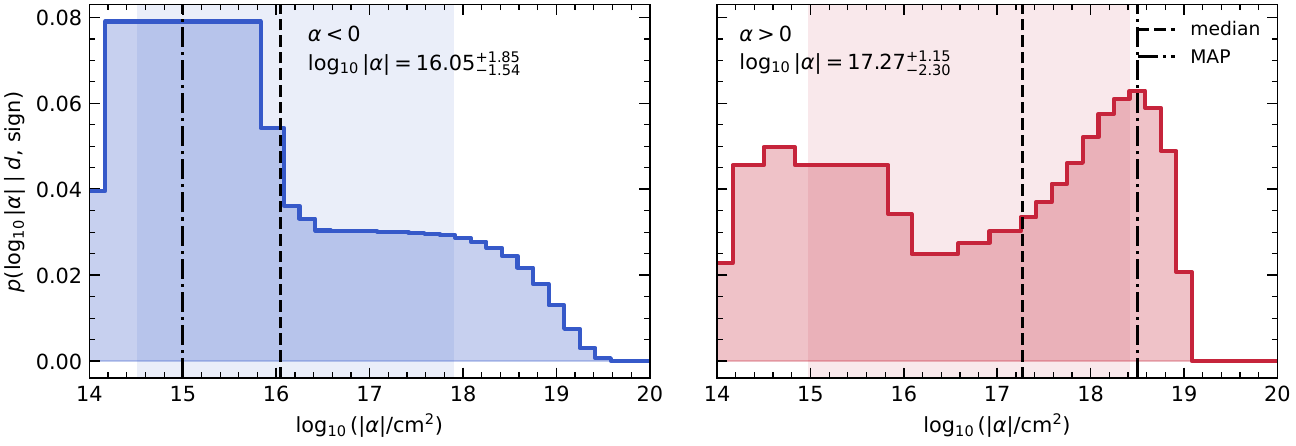}
\caption{\label{fig:posterior-signed}Posterior probability in \(\log_{10}|\alpha|\) conditional on the negative sector (left) and positive sector (right) for the five object sample, ADM mass, and conservative branch. The blue and red areas filled beneath the curves show the corresponding conditional posterior densities. The lighter vertical background bands indicate the central 68 percent credible intervals. Dashed lines mark the posterior medians, while dash dotted lines mark the MAP values. The plotting interval ends at \(\log_{10}|\alpha|=20\); the quadrature and normalization retain the full prior interval through 21. GR is a separate discrete model and is not a bin in either panel.}
\end{figure*}

We next consider the posterior within each continuous sign sector. Figure~\ref{fig:posterior-signed} uses the five object sample, ADM mass, and the conservative branch. The blue and red filled distributions represent the posterior densities conditional on \(\alpha<0\) and \(\alpha>0\), respectively. Both panels use the same prior, uniform in \(\log_{10}(|\alpha|/\mathrm{cm^2})\) on \([14,21]\). Since the conditional posteriors are broad and asymmetric, the posterior median and central 68 percent credible interval provide the primary summaries. The MAP identifies the largest local posterior value and is retained as a secondary diagnostic. Deterministic quadrature is sufficient because each sector contains only one continuous coupling parameter. A converged MCMC calculation with the same likelihood and priors would target the same posterior.

For the negative sector, the conditional result is
\begin{equation}
\begin{aligned}
\alpha_-^{\rm med}&=-1.12\times10^{16}\,\mathrm{cm^2},\\
|\alpha_-|_{68\%}&=
\left[3.22\times10^{14},7.92\times10^{17}\right]\mathrm{cm^2},
\end{aligned}
\label{eq:negative-posterior-summary}
\end{equation}
with \(\alpha_-^{\rm MAP}=-10^{15}\,\mathrm{cm^2}\). For the positive sector,
\begin{equation}
\begin{aligned}
\alpha_+^{\rm med}&=1.85\times10^{17}\,\mathrm{cm^2},\\
\alpha_{+,68\%}&=
\left[9.37\times10^{14},2.61\times10^{18}\right]\mathrm{cm^2},
\end{aligned}
\label{eq:positive-posterior-summary}
\end{equation}
with \(\alpha_+^{\rm MAP}=3.16\times10^{18}\,\mathrm{cm^2}\). These intervals are conditional on the sign, the five object sample, the cold carbon EoS, and the adopted likelihood. Neither interval is a two sided exclusion of GR.

Conditional posteriors do not determine the relative probability of the two signs because each is normalized within its own sector. The global comparison combines their evidences with the discrete GR evidence. Equal prior weights of \(1/3\) for the negative, GR, and positive sectors give the results in Table~\ref{tab:signed-model-comparison}. The positive sector has the largest posterior probability, \(0.395\), but GR follows closely with \(0.378\). The negative sector retains probability \(0.228\). The corresponding log Bayes factors are small on the scale required for model discrimination. In particular, \(\ln B_{+/{\rm GR}}=0.043\), and the evidence ratio between the two continuous sectors is \(\ln B_{-/+}=-0.550\). The five object data therefore do not select a sign or distinguish the quadratic model from GR under the adopted assumptions.

\begin{table}[t]
\caption{Signed model comparison for the five object sample, ADM mass, conservative branch, and equal prior sector weights. Conditional summaries apply only within the corresponding continuous sector.}
\label{tab:signed-model-comparison}
\begin{ruledtabular}
\begin{tabular}{lccc}
Sector & conditional median \((\mathrm{cm^2})\) & \(P(\mathcal M|d)\) & \(\ln B_{\mathcal M/{\rm GR}}\)\\
\colrule
\(\alpha<0\) & \(-1.12\times10^{16}\) & 0.228 & \(-0.507\)\\
GR & not applicable & 0.378 & 0\\
\(\alpha>0\) & \(+1.85\times10^{17}\) & 0.395 & 0.043\\
\end{tabular}
\end{ruledtabular}
\end{table}

\begin{figure}[t]
\includegraphics[width=\columnwidth]{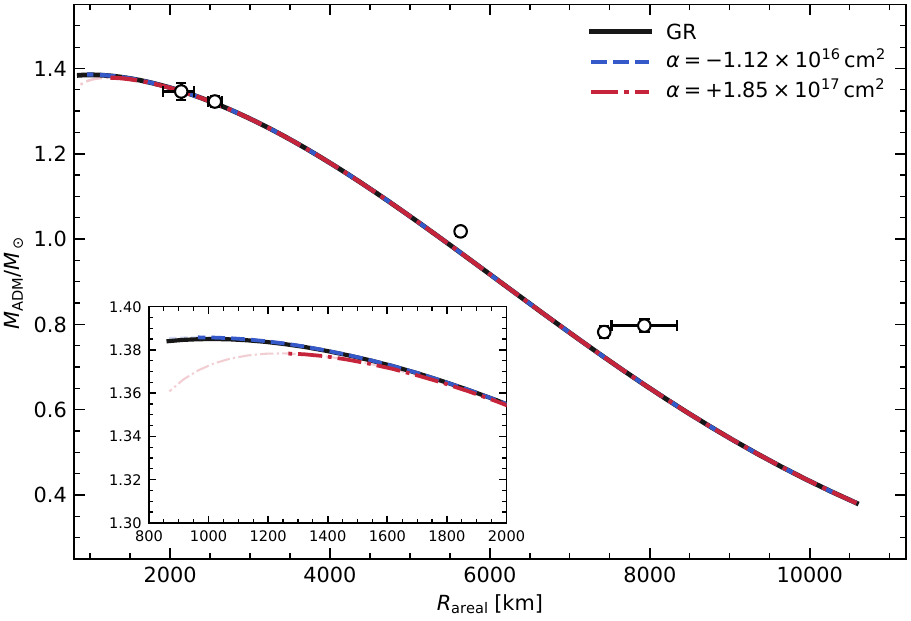}
\caption{\label{fig:median-sequences}Mass and radius sequences calculated at the conditional posterior medians \(\alpha=-1.12\times10^{16}\,\mathrm{cm^2}\) and \(\alpha=+1.85\times10^{17}\,\mathrm{cm^2}\), together with GR and the five observations. The inset enlarges \(800\leq R_{\rm areal}/\mathrm{km}\leq2000\) and \(1.30\leq M_{\rm ADM}/\Msun\leq1.40\). The close curves show that the posterior medians correspond to small observable departures from GR, although larger deviations exist elsewhere in the prior domain.}
\end{figure}

The mass and radius sequences evaluated exactly at the two posterior medians are shown in Fig.~\ref{fig:median-sequences}. They reach equilibrium maxima of \(1.3858\,\Msun\) for the negative median and \(1.3784\,\Msun\) for the positive median, compared with \(1.3850\,\Msun\) in GR. These values describe the calculated equilibrium families and are not stability limits. More importantly, the three curves are nearly coincident throughout most of the observed mass and radius range. The inset resolves their small separation near the compact end. The extreme negative branches allowed by the field equations in Fig.~\ref{fig:mr-family} consequently do not represent the region containing most posterior probability for the five object analysis.

The inference changes when the two ultramassive objects are removed. For Sirius B, QS Vir, and V471 Tau B alone, the sector probabilities are \(P_-=0.332\), \(P_{\rm GR}=0.258\), and \(P_+=0.410\). The conditional medians then become \(\alpha_-^{\rm med}=-6.46\times10^{18}\,\mathrm{cm^2}\) and \(\alpha_+^{\rm med}=3.55\times10^{18}\,\mathrm{cm^2}\). Adding ZTF J1901+1458 and LHS 4033 moves both conditional distributions toward smaller magnitudes and raises the relative probability of GR. Large positive couplings reduce the accessible high mass region, while the ultramassive data favor sequences that remain close to the GR compact branch. This response is physically consistent with Figs.~\ref{fig:mr-observations} and~\ref{fig:mass-density}, but it is also sensitive to applying a static and nonmagnetic model to objects for which those assumptions are incomplete.

The branch definition produces a larger systematic change than the numerical quadrature. For the three direct objects, replacing the conservative branch with the full admitted sequence changes the sector probabilities from \(0.332,0.258,0.410\) to \(0.400,0.275,0.325\) for negative, GR, and positive sectors, respectively. By comparison, varying the nuisance density quadrature changes the log evidences by less than \(8\times10^{-9}\), while thinning the coupling grid changes a log evidence by at most \(0.051\). Leave one out tests also change which sector has the largest probability. The sign ordering is therefore not stable under the main physical and sample choices, even though the quadrature itself is numerically converged.

All posterior statements remain conditional on the cold \(\Ctwelve\) EoS. Rotation, magnetic stresses, finite temperature, composition uncertainty, and observational mass and radius covariance are not marginalized. The present calculation establishes that both signs possess regular ADM matched equilibrium sequences and that their integrated evidences can be compared consistently. It does not provide a detection of modified gravity, a stable high mass negative branch, or a model independent constraint on \(\alpha\).

\section{Conclusions}
\label{sec:conclusions}

We constructed static WD equilibrium sequences for both signs of the quadratic coupling in \(f(Q)=Q+\alpha Q^2\), including GR as the \(\alpha=0\) reference. The regular center expansion selects the affine branch \(C=r e^{B/2}\), and each stellar interior is continued through the vacuum exterior to determine the ADM mass. Combined with the cold \(\Ctwelve\) EoS and the inverse beta density boundary, this construction provides a single setting in which the negative, GR, and positive sectors can be compared from the field equations to the observed mass and radius plane.

The sign of \(\alpha\) controls the high density response of the equilibrium family. Positive couplings shift the compact branch toward larger radii and reduce the mass reached before the geometric conditions terminate the sequence. Negative couplings extend the family toward higher masses and produce the pronounced branches seen at large \(|\alpha|\). The most massive admitted configuration in the explored grid reaches \(14.34\,\Msun\) for \(\alpha=-4.64\times10^{20}\,\mathrm{cm^2}\). This result shows that the negative sector admits a qualitatively distinct hydrostatic-equilibrium regime. Its dynamical behavior, however, requires a separate analysis based on the small radial perturbation equations, which we defer to future work.

The Bayesian analysis uses the same five stars in all three sectors and marginalizes the unobserved central density along each conservative sequence. Within the negative sector, the posterior median is \(\alpha_-^{\rm med}=-1.12\times10^{16}\,\mathrm{cm^2}\), with a central 68 percent interval in magnitude from \(3.22\times10^{14}\) to \(7.92\times10^{17}\,\mathrm{cm^2}\). The positive sector gives \(\alpha_+^{\rm med}=1.85\times10^{17}\,\mathrm{cm^2}\), with a corresponding interval from \(9.37\times10^{14}\) to \(2.61\times10^{18}\,\mathrm{cm^2}\). These median sequences remain close to GR over most of the observed domain, whereas the compact stars probe the region where the theoretical separation between the sign sectors is largest.

With equal prior weights for negative \(\alpha\), GR, and positive \(\alpha\), their posterior probabilities are \(0.228\), \(0.378\), and \(0.395\), respectively. The positive sector receives the largest integrated probability, while its evidence relative to GR is nearly equal, \(\ln B_{+/{\rm GR}}=0.043\). The result is therefore a signed constraint in which the data locate characteristic coupling scales inside each continuous sector and retain substantial probability for all three gravitational descriptions. The change obtained when the two ultramassive stars are removed shows that these compact objects supply much of the sensitivity to the coupling sign.

The exterior mass calculation supports this comparison quantitatively. Over the conservative sequences, the relative difference between the integrated material mass and the ADM mass is at most \(1.44\times10^{-5}\), and replacing one by the other leaves the sector probabilities practically unchanged. Variations of the density quadrature are smaller still, while the choice of admitted equilibrium branch produces the leading change in the inference. The numerical integration is thus sufficiently resolved to expose the physical dependence on the sequence definition and on the stellar sample.

The signed extension changes the physical scope of the model. Negative couplings generate equilibrium configurations that are absent from the positive sector, while the posterior medians favor departures that are much closer to the GR sequence in the mass-radius region currently sampled. Radial oscillations on the selected affine branch could determine the stable dynamical domain of the high mass solutions. Incorporating rotation and magnetic structure for ultramassive WDs, together with alternative compositions, will allow the same signed Bayesian construction to separate gravitational effects from the stellar physics that is most relevant at the compact end.

\appendix

\section{Reproducibility and resolution summary}
\label{app:reproducibility}

The signed sequence set contains 75 values of the coupling: 37 negative values, GR, and 37 positive values. The magnitude grid covers \(10^{14}\leq|\alpha|/\mathrm{cm^2}\leq10^{21}\) and contains 4529 accepted stellar configurations. Each configuration is tested for center regularity, the radial constraint, \(f_Q>0\), monotonicity of \(C(r)\), exterior convergence, and the equation of state density boundary. No likelihood point is interpolated through a rejected interval or between disconnected branches. The observable pair is \([\MADM,C(r_s)]\), and the conservative mask ends at the first maximum of \(M(\rho_c)\).

The stellar calculation was repeated for 18 signed test configurations under tighter numerical settings. The largest relative changes were \(6.33\times10^{-5}\) in \(\MADM\) and \(9.17\times10^{-5}\) in \(C(r_s)\). The maximum normalized angular equation closure residual was \(1.76\times10^{-7}\). Near the GR limit, the largest relative difference in the direct continuity check was \(4.90\times10^{-6}\). These tests complement the symbolic GR and fixed \(C\) limits, the independent GR sequence, the residual profiles, and the exterior tail convergence calculation.

For each sign sector, the evidence is integrated with trapezoidal weights in \(x=\log_{10}(|\alpha|/\mathrm{cm^2})\). The nuisance central density integral uses 4097 points with normalized trapezoidal weights in \(\ln\rho_c\). Refining this quadrature changed a log evidence by at most \(7.31\times10^{-9}\) and a sector probability by at most \(2.27\times10^{-9}\). Thinning the signed coupling grid changed a log evidence by at most \(0.051\). The mass definition replacement, branch selection tests, sample variations, and prior weight sensitivity were evaluated independently and retained as separate products of the signed analysis.

\begin{acknowledgments}
EO thanks the Fundação Cearense de Apoio ao Desenvolvimento Científico e Tecnológico (FUNCAP), through grant BP6-0241-00335.01.00/25.
JMZP acknowledges the financial support provided by FAPERJ under Process No.~SEI-260003/000308/2024.
\end{acknowledgments}

\vspace{0.5cm}
\section*{Data availability}

The sequence, likelihood, posterior, domain, and validation products used in this manuscript are stored in the accompanying project workspace. A public repository identifier is to be confirmed before submission.

\bibliographystyle{apsrev4-2}
\bibliography{references}

\end{document}